\documentclass[graybox]{svmult}

\usepackage{mathptmx}       
\usepackage{helvet}         
\usepackage{courier}        
\usepackage{type1cm}        
\usepackage{makeidx}         
\usepackage{graphicx}        
\usepackage{multicol}        
\usepackage[bottom]{footmisc}

\makeindex             

\begin{document}

\title*{Stellar activity cycles and contribution of the deep layers knowledge}
\author{S. Mathur}
\institute{S. Mathur \at High Altitude Observatory, NCAR, P.O. Box 3000, Boulder, CO 80307, USA, \email{savita@ucar.edu}}
%
%
\maketitle
\abstract*{It is believed that magnetic activity on the Sun and solar-type stars are tightly related to the dynamo process driven by the interaction between rotation, convection, and magnetic field. However, the detailed mechanisms of this process are still incompletely understood. Many questions remain unanswered, e.g.: why some stars are more active than others?; why some stars have a flat activity?; why is there a Maunder minimum?; are all the cycles regular?
A large number of proxies are typically used to study the magnetic activity of stars as we cannot resolve stellar discs.
Recently, it was shown that asteroseismology can also be used to study stellar activity, making it an even more powerful tool. If short cycles are not so uncommon, we expect to detect many of them with missions such as CoRoT, {\it Kepler}, and possibly the PLATO mission. 
We will review some of the latest results obtained with spectroscopic measurements. We will show how asteroseismology can help us to better understand the complex process of dynamo and illustrate how the CoRoT and {\it Kepler} missions are revolutionizing our knowledge on stellar activity. A new window is being opened over our understanding of the magnetic variability of stars.}

\abstract{It is believed that magnetic activity on the Sun and solar-type stars are tightly related to the dynamo process driven by the interaction between rotation, convection, and magnetic field. However, the detailed mechanisms of this process are still incompletely understood. Many questions remain unanswered, e.g.: why some stars are more active than others?; why some stars have a flat activity?; why is there a Maunder minimum?; are all the cycles regular?
A large number of proxies are typically used to study the magnetic activity of stars as we cannot resolve stellar discs.
Recently, it was shown that asteroseismology can also be used to study stellar activity, making it an even more powerful tool. If short cycles are not so uncommon, we expect to detect many of them with missions such as CoRoT, {\it Kepler}, and possibly the PLATO mission. 
We will review some of the latest results obtained with spectroscopic measurements. We will show how asteroseismology can help us to better understand the complex process of dynamo and illustrate how the CoRoT and {\it Kepler} missions are revolutionizing our knowledge on stellar activity. A new window is being opened over our understanding of the magnetic variability of stars.}

\section{Why study stellar magnetic activity?}
\label{sec:1}

The study of magnetic cycles in stars other than the Sun is important  from the point of view of {\it solar-stellar connexion}. The detailed mechanisms of the dynamo are not yet well known as we still have difficulties in predicting the solar cycles. An example is the solar minimum between cycles 23 and 24 that was unexpectedly long (e.g. \cite{How09, Sal09}). We know that the dynamo results from the interaction between rotation (latitudinal differential rotation), convection, and magnetic field. By studying stellar cycles, we can expect to better understand how dynamos operate under several physical conditions (rotation rate, depth of convective zone), assuming that dynamo operates similarly in solar-type stars compared to the Sun. This work would put additional constraints on solar dynamos models, for which, there is a variety of scenarios: the mean-field model, the thin-shell model, and the distributed dynamos (e.g. \cite{Cho95, MacG97, Dik05, Jou08}).  As the dynamo takes place in the outer layers of the stars, it is required to have a good knowledge on the structure of the stars. For instance, the position of the base of the convective zone is a crucial ingredient and it seems that the tachocline plays a role in the efficiency of the dynamo but its exact contribution remains unclear. Over the past decade, asteroseismology has emerged as a powerful unique tool allowing us to directly probe the stellar interior. Aside from determining the radii, masses, and ages of stars more precisely than with classical techniques \cite{Met10}, with asteroseismology we can try to fit stellar models with observed frequencies \cite{Mat11s} and use acoustic glitches \cite{Bal04} to have estimates of the depth of the convection zone of the stars. Beside, using photometric observations, people developed or applied different techniques to study the surface rotation, such as spot modeling, interferometric imaging or wavelet techniques. The detection of mixed modes (\cite{Bec11, Bed11, Cam11, Mat11b}) can also provide information on the rotation of the stellar interior, specially the core, as these modes are the result of the coupling between acoustic (p) and gravity (g) modes. The NASA mission {\it Kepler} \cite{Bor10} has already made a step further concerning this topic by unveiling the radial differential rotation in a few red-giant stars \cite{Bec11b}. By combining this information on the rotation profile of a star and its magnetic activity, we should have more constraints on the stellar dynamo models (e.g. \cite{Bro11}).

Another impact of the stellar activity study would be in the field of the {\it star-planet interaction}. Within a planetary system, stars are the largest source of energy \cite{Rib10}. The emission of particles, through winds for instance, can affect the composition, thermal properties and the atmosphere of the planet(s). The existence and the measurement of the strength of stellar winds were studied with H Lyman $\alpha$ lines \cite{Woo02} These winds can be responsible for the increase of atmospheric temperature of the planet and erosions on the surface of the planet. These particle emissions can also trigger some photochemical reactions (e.g. \cite{Bau04}). Observations of X-rays and far UV (FUV) for several stars with different ages were analyzed by \cite{Rib05} who showed that the older the star gets, the lower the X-ray and FUV fluxes become. These radiations are important to study the development of life on the planets and bring another factor to take into account for the definition of the habitable zones of exoplanets. Finally, there is a debate about the impact of the planet on the activity of host stars. Some tidal and magnetic scenarios have been developed to explain these phenomena while \cite{Sho05}, for example, suggested that planets could trigger a release of energy that was built up in coronal loops.  Several teams also tried to study if close-in giant planets can have an impact on host-stars X-ray emission but so far no clear signature has been found (\cite{Kas08, Sch10, Pop11}.

\section{Classical observations}
\label{sec:2}

A large number of proxies are typically used to study the magnetic activity of stars, such as optical, UV or X-ray emission from magnetically heated gas but also starspots or flares, and other phenomena that are associated with magnetic fields in the Sun. 
Many surveys have been carried out to measure UV, X-ray and CaHK in the stellar atmosphere. The common point between all of them is that they depend on the surface activity and possibly on the inclination angle of the star and the position of the active latitudes. In this section, we will present  the results that have been obtained on stellar cycles thanks to these proxies. 

\subsection{The Mount Wilson project and some results}


One of the most used proxy is the Mount Wilson S-index, which is sensitive to the photosphere and the chromosphere. It requires a high contrast between the atmospheric energy and the underlying thermal spectrum, making ground-based observations possible for these kinds of observations. It is thus defined as the ratio between the flux in HK lines and the flux in R and V bandpasses \cite{Wil78}. The average value for the Sun is around 0.171. Another proxy, $\log (R'_{\rm HK})$, is sensitive to the chromosphere. It is defined as the ratio between the emission from chromosphere in CaHK and the total bolometric emission of the star. The typical average value for the Sun is of -4.9.

In the 60s a huge project of CaHK observations started to study the temporal variation of chromospheric emission in a large amount of main-sequence stars. This project was led at the Mount Wilson Observatory, leading to 40 years of data. 

From a sample of $\sim$\,110 stars monitored by the project, \cite{Bal95} showed that the activity level, the rotation period, the presence of a Maunder minimum, the regularity of the cycles could vary with the evolutionary stage of the star. They also found that 60\% of lower MS stars are similar to the Sun, 25\% of stars seem to be variable, and 15\% of the stars seem to have a flat activity.

By analyzing another sample of stars monitored by this project, \cite{Rad98} suggested that the total stellar irradiance varies  with chromospheric activity. In particular, young and more active stars appear to be fainter when HK emission increases while old and less active stars (like Sun) are brighter when HK emission increases. This is in agreement with what we observe with the Sun where the brightness increases when more sunspots are present on its surface.

Some studies were also done to find a correlation between the cycle period and the rotation period of the star. For cool stars like the Sun, with an $\alpha \Omega$ dynamo, the longer the period of rotation is, the longer the cycle period will be \cite{Saa02}.
Indeed, an empirical formula has been derived (e.g. \cite{Jou10}) where $P_{\rm cyc}/P_{\rm rot}$ depends on the convective turnover time. But due to large uncertainties on some parameters, it is difficult to have a precise estimate of $P_{\rm cyc}$. Finally, \cite{Boh07} studied some of the Wilson project stars and found a very interesting result showing that two different branches seem to exist among the stars: the active and inactive branches. No clear explanation has been given so far but some possible scenarios involve the $\alpha$ effect or maybe different locations of the dynamo shell.

\subsection{Some interesting cases}


{\bf A solar twin, 18 Sco:} This solar twin has a mass and a radius very close to the ones of the Sun \cite{Baz11}. \cite{Pet08} estimated its rotation period to 22.7$\pm 0.5$~days. Using $\sim$~10 years of spectroscopic and photometric data from the Solar-Stellar Spectrograph and the Automatic Photospheric Telescope, an excess flux was found in the CaHK measurements \cite{Hal00}. Beside, simultaneous measurements in b and y passbands were done and they presented similar variations to the Sun. The average Mount Wilson S-index was estimated to $\sim$~0.182 and the cycle period $P_{\rm cyc} \sim$~7~yrs.  



\noindent {\bf Short activity cycles:} The star $\iota$~Hor (or HD170151) is known to be hosting an exoplanet. It has been observed by the Cero Tololo Observatory in Chile for 2.5~years.  The Mount Wilson S-index of these data varied from 0.21 to 0.28, which is around 30\% more than what we see in the Sun. The Lomb-Scargle periodogram suggests a rotation period of 8 days, which means that this is a rather fast rotator. \cite{Met10b} fitted the data and found a cycle period of 1.6 years, putting this star in the category of stars with the shortest cycle periods detected. This result is very promising as it suggests that short cycles are more common than we thought and with the present missions, we could expect to observe some additional magnetic activity cycles in other stars.


\noindent {\bf Magnetic field reversal:} Today, it is possible to study the magnetic field topology of stars using spectropolarimetric measurements with instruments such as ESPaDOnS or NARVAL. \cite{Pet09} studied the evolution of the large-scale photospheric magnetic field geometry of the solar-type star HD 190771 with NARVAL data. This star is a solar analog as well but with a rotation period of $\sim$~8 days. Their analysis put in evidence a polarity reversal between observations separated by 1 year as well as dramatic change of the toroidal field over a period of a few months. This kind of information is very important to understand the impact of the rotation on the generation of magnetic fields through the dynamo mechanism.

\section{A new era with seismology}
\label{sec:3}

The new generation of space missions such as CoRoT \cite{Bag06} and {\it Kepler} are allowing us to do asteroseismology by providing exquisite photometric data (e.g. \cite{Cha11sci}). Among the solar-type stars, many light curves present some variability that can be attributed to the presence of starspots crossing the stellar surface and thus suggesting that these stars are magnetically active. One way of studying the magnetic activity of these stars (apart from using classical observations as in Section~\ref{sec:2}), we can use spot modeling (\cite{Dor87, Lan03}), which has already given results with some CoRoT and {\it Kepler} targets (\cite{Lan09, Mos09, Mat10corot, Bal11, Fra11}). This modeling is based on a large number of parameters but can retrieve estimates of the lifetime of the spots, the surface rotation, and the presence of differential rotation.

Beside, we also know that for the Sun, when the magnetic activity increases, the frequency of the p modes increases while their amplitude decreases \cite{Sal09}. The reason why we observe these changes are still under investigation but they are probably related to changes in the outer layers of the stars that affect the outer turner point of the modes. Aside that, it also seems that though no surface magnetic activity is observed (such as a Maunder minimum) a possible cycle could still be running underneath the photosphere. Prediction of the amplitude of frequency shifts using 2 different scalings between the frequency shift and the variation of the $R'_{\rm HK}$ index have been done. While \cite{Cha07} predicted higher amplitudes for G stars, \cite{Met07} predicted higher amplitudes for F stars.
Recently, \cite{Gar10} discovered the first signature of magnetic activity using seismology in an F star: HD~49933, agreeing with the predictions by \cite{Met07}. CaHK observations in Chile confirmed that this star is magnetically active with a cycle period of at least 210 days. A deeper analysis of the frequency shifts of the p modes as a function of the frequency reveals that the high-frequency modes are more shifted than the low-frequency modes, which is similar to what we observe in the Sun \cite{Sal11}, suggesting that these frequency shifts are related to magnetic activity occurring in the outer layers of the star. A similar analysis has been done for three other CoRoT targets \cite{Mat11b} that were also observed by the spectropolarimeter NARVAL. There is some hint of activity but more data would be necessary to confirm any clear detection. An ensemble analysis of hundreds of solar-type stars observed by {\it Kepler} also showed that stars that seem to be more active have lower p-mode amplitude and even undetectable modes \cite{Cha11b}. 

We are starting to have many observations of magnetic activity and cycles in other stars while asteroseismology is also providing information on the stellar interiors. It is just a question of time to combine both information to improve our understanding of dynamo process and how the magnetic field evolves in the outer layers of the stars.

\begin{acknowledgement}

NCAR is supported by the National Science Foundation. This research was supported in part by the National Science Foundation under Grant No. NSF PHY05-51164.
\end{acknowledgement}

\end{document}